# Experimental signature of parity anomaly in semi-magnetic topological insulator

M. Mogi[1†*], Y. Okamura[1], M. Kawamura[2], R. Yoshimi[2], K. Yasuda[1†],

A. Tsukazaki[3], K. S. Takahashi[2], T. Morimoto[1], N. Nagaosa[1,2],

M. Kawasaki[1,2], Y. Takahashi[1,2], and Y. Tokura[1,2,4]*

[1] *Department of Applied Physics and Quantum Phase Electronics Center (QPEC), University of Tokyo, Bunkyo-ku, Tokyo 113-8656, Japan.*

[2] *RIKEN Center for Emergent Matter Science (CEMS),*

*Wako, Saitama 351-0198, Japan.*

[3] *Institute for Materials Research, Tohoku University,*

*Sendai, Miyagi 980-8577, Japan.*

[4] *Tokyo College, University of Tokyo, Bunkyo-ku, Tokyo 113-8656, Japan.*

[†] *Present address: Department of Physics, Massachusetts Institute of Technology, Cambridge, MA 02139, USA.*

*e-mail: mogi@mit.edu; tokura@riken.jp

**A three-dimensional topological insulator features a two-dimensional surface state consisting of a single linearly-dispersive Dirac cone[1-3]. Under broken time-reversal symmetry, the single Dirac cone is predicted to cause half-integer quantization of Hall conductance, which is a manifestation of the parity anomaly in quantum field theory[1-8]. However, despite various observations of quantization phenomena[9-14], the half-integer quantization has been elusive because a pair of equivalent Dirac cones[15] on two opposing surfaces are simultaneously measured in ordinary experiments. Here we demonstrate the half-integer quantization of Hall conductance in a synthetic heterostructure termed a 'semi-magnetic' topological insulator, where only one surface state is gapped by magnetic doping and the opposite one is non-magnetic and gapless. We observe half quantized Faraday/Kerr rotations with terahertz magneto-optical spectroscopy and half quantized Hall conductance in transport at zero magnetic field. Our results suggest a condensed-matter realization of the parity anomaly[4-8] and open a way for studying unconventional physics enabled by a single Dirac fermion.**

Realization of relativistic quantum phenomena is one of the central issues of contemporary condensed-matter physics. A prominent example is the emergence of Dirac fermions in two-dimensional (2D) systems[16], which are characterized by a linearly dispersive electronic band called a Dirac cone with spin-momentum locking and a crossing point (Dirac point) at zero energy. When time-reversal symmetry is broken, an energy gap opens at the Dirac point. Focusing on the spin degrees of freedom in a single Dirac cone, it turns out that this gap opening is simultaneously accompanied by 2D parity (equivalent to mirror) symmetry breaking, which is known as the parity anomaly in quantum field theory[4-8]. A notable consequence of the parity anomaly is the generation of a parity-violating current with the half-integer quantization of Hall conductance, $e^2/(2h)$, under an electromagnetic field perturbation, where $e$ is the elementary charge and $h$ is Planck's constant. However, according to Nielsen-Ninomiya's fermion doubling theorem[15], the Dirac cones always appear in pairs and the paired Dirac cones cancel the parity anomaly, restoring the parity symmetry as a whole system. Therefore, the quantization phenomena reported so far, such as the quantum Hall effect[16], valley Hall effect[17], and quantized magneto-optical effects[18,19], are characterized by integer topological indices[20]. In the meanwhile, the possibility of condensed-matter realization of the parity anomaly associated with two Dirac cones was proposed by Haldane in a honeycomb lattice[8], which was realized in a cold atom system[21]. However, the realization of the single Dirac cone and the half-integer quantization of Hall conductance has been elusive.

Here, we demonstrate that the half-integer quantization of Hall conductance associated with the parity anomaly can be realized in three-dimensional (3D) topological insulators (TIs)[1-3]. The 3D TIs support a single gapless Dirac fermion cone at each surface as a consequence of the nontrivial $Z_2$ topological nature of wavefunctions in the insulating bulk. Unlike the 2D lattice systems where the paired Dirac cones, degenerated by spin and orbital degrees of freedom, exist at different points of momentum space, a pair of spin-polarized single Dirac cones in the 3D TI appear separately on the opposite surfaces, specifically top and bottom surfaces in a crystal of thin-film form. Over a decade ago, the potential half-integer quantization at each surface was firstly recognized by Fu and Kane[2], then theoretically demonstrated in the layer-resolved Hall conductivity calculation by Essin et al.[22] In this experimental study, we utilize a layer-resolved doping technique of magnetic ions targeting only the vicinity of the top surface (Fig. 1a). It is possible to gap out only the top-surface Dirac cone with simultaneously broken 2D parity symmetry and time-reversal symmetry while keeping the bottom-surface Dirac cone gapless[23,24] (Fig. 1b); that breaks the inversion symmetry as a 3D structure (Supplementary Information I). Since only one of the paired Dirac cones is gapped as well as the 3D bulk part is also insulating, this 'semi-magnetic' TI would be an ideal arena for the demonstration of the parity anomaly.

The semi-magnetic TI heterostructure films consisting of (Bi, Sb)$_2$Te$_3$ and Cr-doped (Bi, Sb)$_2$Te$_3$ (refs. 9, 11) were grown by molecular-beam epitaxy (Fig. 1c) (see Methods and Supplementary Information II). The magnetic element Cr was modulation-doped only near the top surface (2 nm). The exchange interaction between the surface electrons and the magnetic ions (Cr) then opens an energy gap on the top surface Dirac cone. The Fermi energy $E_F$ was carefully tuned so that $E_F$ lies within the magnetic gap of the top surface Dirac cone, confirmed by the gate-voltage-dependent transport measurements (see Supplementary Information V). The energy level of the Dirac point for the bottom surface ($E_{DP}$) remains tunable by changing the Bi:Sb ratio ($x$).

One of the experimental methods to observe the half-integer quantization is terahertz (THz) magneto-optical Faraday and Kerr rotation measurements[3]. Previous studies reported the integer quantization of Faraday and Kerr rotations in uniform TI films with broken time-reversal symmetry[12-14]. In these experiments, the top and bottom surface states simultaneously contribute to the quantized rotations because the magneto-optical polarization rotations integrate all the rotatory contributions along the light path[18,25,26]. By contrast, in the case of the semi-magnetic TI, it is expected to show the half-quantized magneto-optical rotations because only the gapped top surface can be a source of the magneto-optical rotations.

We first present the Faraday and Kerr rotation spectra for the magnetic TI film which exhibits the quantum anomalous Hall (QAH) effect; the spectra are shown in Figs. 2c and 2d as a function of the incident light ($\hbar\omega$, where $\hbar = h/2\pi$) taken at zero magnetic field $\mu_0 H = 0$ T and temperature $T = 1$ K once after training the magnetization up to 2 T, which was obtained by analyzing the time-domain waveforms of the transmitted THz pulses (Figs. 2a and 2b)[27,28]. We find that the real parts of Faraday $\theta_F$ and Kerr $\theta_K$ rotation angles with the negligible imaginary parts (ellipticities) of $\eta_F$ and $\eta_K$ are consistent with the theoretical values of the quantized Faraday and Kerr rotations: $\tan^{-1}\left(\frac{2\alpha}{n_s+1}\right) = 3.27$ mrad and $\tan^{-1}\left(\frac{4n_s\alpha}{n_s^2-1}\right) = 9.20$ mrad, respectively, where $n_s \sim 3.46$ is the refractive index of InP substrates, $\alpha = \frac{e^2}{2\varepsilon_0 hc} \approx \frac{1}{137}$ is the fine structure constant, $\varepsilon_0$ is the permittivity of the vacuum, and $c$ is the speed of light. The result is in accord with the previously reported result[14] while the signal-to-noise ratio is much improved in this measurement (see Extended Data Fig. 1).

We now present the result of the magneto-optical rotations for a semi-magnetic TI film with the use of the same experimental setup. As shown in Fig. 2E, we observed that $\theta_F$ takes around 1.6 mrad, being almost independent of the incident light frequency. This value is almost half of that for the QAH (Fig. 2c) and integer ($\nu = 1$) quantum Hall (QH) systems[12-14,18,19,25,26]. This half Faraday rotation angle is accounted for by the half-integer quantization in the magneto-optical rotation: The Faraday rotation, for small rotation angles, is described by the summation of the dynamical 2D Hall conductivity $\sigma_{xy}^{2D}(\omega, z)$ along the light path ($z$-direction), $\tan\theta_F = \frac{1}{(n_s+1)c\varepsilon_0} \sum_i \sigma_{xy}^{2D}(\omega, z_i)$ ($z_i$: $z$ coordinate of the $i$-th layer)[18]. Given that the top and bottom surface states are well decoupled and that the gapped top surface state has half quantized Hall conductivity while the gapless bottom surface and bulk states have no contribution, the rotation angle should be $\theta_F = \tan^{-1}\left(\frac{\alpha}{n_s+1}\right) \approx 1.63$ mrad, being consistent with the experimentally measured values around 1.6 mrad.

Similarly, the Kerr rotation angle $\theta_K$ was measured by utilizing multiple reflections of the THz pulses inside the substrate (Fig. 2b)[27,28]. As shown in Fig. 2f, the value of $\theta_K$ is around 4.5 mrad, which is again almost half of that for the QAH state (Fig. 2d). This also agrees with the predicted half-integer quantization for a thin film limit ($\theta_K = \tan^{-1}\left(\frac{2n_s\alpha}{n_s^2-1}\right) \approx 4.60$ mrad) (Supplementary Information III). The ellipticities $\eta_F$ and $\eta_K$ are quite small compared with $\theta_F$ and $\theta_K$, respectively, suggesting no resonance feature due to a sufficient magnetic gap opening of the top surface state compared to the incident THz light energy ($\hbar\omega < 5$ meV) (Supplementary Information IV). Thus, the observation of the half quantized rotations under $\mu_0H = 0$ T is consistent with the picture that only the gapped top surface works as a source of magneto-optical rotations and the contribution of the gapless bottom surface to the rotations is negligibly small (Supplementary Information III).

The half-quantized magneto-optical rotations at the low-frequency region motivated us to study the Hall conductance at a zero-frequency limit, namely dc electrical transport measurements. In the magneto-optical rotation measurements, the observed Faraday and Kerr rotations can be regarded as a consequence of the topological nature in the interior of the gapped top surface. In dc electrical transport measurements, on the other hand, only the gapless bottom surface state where the electric current is running through is to be probed. Thus, confirming the correspondence between the magneto-optical rotation and dc electrical transport measurements is important for understanding the surface-bulk correspondence of the low-energy half-integer quantized electrodynamics in TI. Experimental efforts to probe it have been made in similar magnetic/non-magnetic TI thin flakes under high magnetic fields[29,30], which however should result in the integer quantization or taking arbitrary non-integer quantized values due to both contributions of top and bottom surface states (Supplementary Information IX). Seriously, the constraints of flake shapes with invasive metal-contacts cannot avoid the occurrence of artifacts in the Hall conductance converted from the resistance due to inevitable non-uniform current densities and equipotential lines in the samples[31].

In Figs. 3a and 3b, we show the temperature ($T$) dependence of the Hall resistance ($\rho_{yx}$) and the sheet resistance ($\rho_{xx}$) at zero magnetic field, respectively, for five Hall bar devices with the well-defined sample dimensions by photolithography for accurate Hall measurements (inset of Fig. 3d and Extended Data Fig. 3). Bi:Sb ratios ($x$) correspond to different $E_{DP}$ (the energy level of the Dirac point for the bottom surface state); see the inset of Fig. 3a and Fig. 1b. Apparently, there are no characteristic features common to the respective samples. By contrast, when resistance is converted to conductance, the values of $\sigma_{xy}$ for all the samples converge to the half quantum conductance $0.5e^2/h$ at low temperatures as clearly seen in Fig. 3c. The quantization occurs below 2 K, which is comparable to the typical quantization temperature of the QAH effect (Figs. 3a and 3b). This result indicates that regardless of different $E_{DP}$, $\sigma_{xy}$ is quantized to the half-integer value as long as $E_F$ resides in the magnetic gap of the top surface state. Figure 3D shows the data of the sheet conductance $\sigma_{xx}$ which shows a relatively large variation with $x$. $\sigma_{xx}$ can be attributed to the dissipative electron transport in the bottom surface where the carrier densities are changed by $x$ (Supplementary Information Figs. S5 and S6).

In Fig. 4a, we directly compare the results of the magneto-optical rotation and the dc electric transport measurements using an identical semi-magnetic TI sample ($x = 0.93$) by using the relation of $\tan\theta_K = \frac{1}{c\varepsilon_0}\frac{2n_s}{n_s^2-1}\sigma_{xy}$ (refs. 18, 25, 26). The results of magneto-optics and transport show a good agreement with each other, meaning that the half-integer quantization is verified regardless of measurement methods. When we apply a perpendicular magnetic field, both $\sigma_{xy}$ and $\theta_K$ are almost doubled and exhibit the integer quantization[11]: $\sigma_{xy} = e^2/h$ and $\theta_K = \tan^{-1}\left(\frac{4n_s\alpha}{n_s^2-1}\right) = 9.20$ mrad (see also Extended Data Fig. 2). The doubled rotations under strong magnetic fields can be understood by the Landau level formation on the bottom surface which works as an additional source of the half-quantized rotations.

The half-integer quantization in transport can be explained by the summation of the half-quantized 2D Hall conductivity ($\sigma_{xy}^{2D,t} = \frac{e^2}{2h}$) with zero 2D longitudinal conductivity ($\sigma_{xx}^{2D,t} = 0$) for the top surface and finite 2D conductivity ($\sigma_{xx}^{2D,b} \neq 0$) with zero 2D Hall conductivity ($\sigma_{xy}^{2D,b} = 0$) for the bottom surface (Supplementary Information Figs. S10) by assuming surface parallel conduction channels, where the superscript t (b) denotes the top (bottom) surface. To provide more insight into the transport, we sketch the current distribution for the top and bottom surfaces (Figs. 4b and 4c). In this picture, the electric current $I = (j_x^t + j_x^b)W$ ($W$: the sample width) is driven along the $x$-direction (Fig. 4c). The dissipative bottom surface current density $j_x^b$ generates the longitudinal electric field as described by $E_x = j_x^b/\sigma_{xx}^b$. Since the top and bottom surfaces are electrically shorted at the side surface, this $E_x$ also generates the non-dissipative top surface Hall current $j_y^t = \sigma_{yx}^{2D,t} E_x = -\frac{e^2}{2h} E_x$ because the top surface state is gapped. Importantly, because the y-direction current densities on the top and bottom surfaces must compensate for each other, $j_y^t + j_y^b = 0$ (Fig. 4c). This additional dissipative current density $j_y^b$ generates the transverse electric field $E_y = j_y^b/\sigma_{xx}^{2D,b}$ which induces the *longitudinal* Hall current $j_x^t = \sigma_{xy}^{2D,t} E_y = \frac{e^2}{2h} E_y$ on the top surface. Eventually, since the experimentally measured resistance components are expressed as $\rho_{xx}^{meas} = E_x/(I/W)$ and $\rho_{yx}^{meas} = E_y/(I/W)$, the Hall conductance becomes $\sigma_{xy}^{meas} = \frac{\rho_{yx}^{meas}}{(\rho_{xx}^{meas})^2+(\rho_{yx}^{meas})^2} = \sigma_{xy}^{2D,t} = \frac{e^2}{2h}$. Thus, $\sigma_{xy}^{meas}$ exhibits the half-integer quantization in the transport measurement irrespective of $\sigma_{xx}^{2D,b}$, which well describes the observations shown in Figs. 3c and 3d. While the current distribution can be comprehensively understood by the above parallel surface conduction picture, the Hall current from the parity anomaly flows through the delocalized gapless states in the side and bottom surfaces in the real space picture (Supplementary Information VII). This situation is analogous to the plateau to plateau transition in the ordinary QH systems, where the delocalized states in the bulk support the Hall current[32]. Here the gapless Dirac fermion in the bottom surface plays the role of such delocalized states (Supplementary Information IX).

Finally, we mention the precision of the half quantized $\sigma_{xy}$. Experimentally, the measured deviation of $\sigma_{xy}$ from $e^2/(2h)$ was about 2.6 % (inset of Fig. 3c). Because $\sigma_{xx}$ remains finite, the precision of the half-integer quantization cannot be so high as the QAH effect, in principle. Furthermore, in the ground state, namely $T \to 0$ K, all the electronic states would be localized as seen in the decrease of $\sigma_{xx}$ with lowering the temperature (Fig. 3d) due to broken time-reversal symmetry as a whole system[33], such as a tiny amount of magnetic (Cr) impurities possibly contained in the bottom surface (Supplementary Information VIII) or else due to a tiny energy gap formed on the bottom surface by top-bottom surface hybridization (Supplementary Information X). This would cause a shift of $\sigma_{xy}$ to either $e^2/h$ or 0, respectively. Nevertheless, the experimentally observed half-integer quantization has surprisingly high stability in the parameter ranges of the present experiments varying $E_F$ (Fig. 3, Supplementary Information Fig. S9), $T$ (down to 50 mK), and the sample size (10 μm to 1.5 mm) (Supplementary Information XI).

In summary, we have experimentally demonstrated the parity anomaly with the half-integer quantization of Hall conductance in the semi-magnetic TI with two different methods of bulk-sensitive magneto-optical spectroscopy and surface-sensitive transport measurements. The semi-magnetic TI structure with the parity anomaly can be a platform for exotic single Dirac fermion physics. For instance, dyon particles arise from the fractional topological magneto-electric effect on the magnetically gapped (top) surface[3,34]. Besides, the additional superconducting proximity coupling on the gapless (bottom) surface can open a superconducting gap. This potentially produces more accessible non-Abelian Majorana edge states under relaxed conditions of magnetic and superconducting gaps[35-38].

**Methods**

**Thin-film growth.** The Cr-doped (Bi, Sb)$_2$Te$_3$ heterostructure films were grown on epi-ready semi-insulating (>10$^7$ Ω cm) InP(111)A substrates maintained at 200 °C (growth temperature) employing a molecular-beam epitaxy (MBE) chamber equipped with standard Knudsen cells under a back pressure of about 1×10$^{-7}$ Pa. The composition of the films was nominally determined with the beam equivalent pressures of Cr, Bi and Sb fluxes, and the beam pressure of Te relative to Bi and Sb was kept at a ratio of about 40:1 to suppress Te vacancies. The Cr modulation-doping was controlled by opening and closing the shutter of the Cr cell. By cross-sectional scanning energy dispersive x-ray spectrometry presented in the previous study[11], where the growth condition including the growth temperature (200 °C) is the same as that in the present study, interlayer diffusion of Cr atoms from the Cr-modulation doped layer is negligibly small. The growth rate of the films was ~ 0.2 nm min$^{-1}$ as characterized by x-ray reflectivity in representative samples. To prevent deterioration of the films, they were capped with AlO$_x$ (~ 5 nm) deposited by an atomic layer deposition (ALD) system at room temperature immediately after taking them out from the MBE chamber.

**Magneto-optical THz spectroscopy.** In the time-domain THz measurements, a mode-locked Ti: sapphire laser generated laser pulses with a wavelength of 800 nm and a duration of 100 fs, which were split into two paths to emit and detect THz pulses by a photoconductive antenna and by a dipole antenna, respectively. The THz photon energy ($\hbar\omega$) range is 1 to 6 meV (0.2 to 1.5 THz in frequency) and is centered at ~ 2 meV. The film samples (~ 1 cm × 1 cm in area) were mounted on a copper stage with a 5 mm hole in diameter for sufficient transmission of THz light and were cooled in a cryostat equipped with a superconducting magnet (7 T) and a $^3$He probe insert down to 1 K. We measured transmitted THz electric fields $E_x(t)$ and $E_y(t)$ using two wire grid polarizers (parallel for the $E_x$ measurements or orthogonal for the $E_y$ measurements) across the samples. To eliminate the background signal except for the magneto-optical rotations, $E_y(t)$ was deduced by anti-symmetrizing the waveforms of $E_y(+H)$ and $E_y(-H)$ ($E_y = (E_y(+H) - E_y(-H))/2$), where $H$ is the magnetic field applied perpendicular to the film planes. When the measurements were done at $H = 0$, the anti-symmetrizing procedure was performed for $E_y(+M)$ and $E_y(-M)$, by reversing the spontaneous magnetization $M$ along $z$ (film normal) direction. The energy spectra of the complex rotation angles are obtained by the Fourier analysis of $E_y(\omega)/E_x(\omega) = (\sin\theta(\omega) + i\eta(\omega)\cos\theta(\omega))/(\cos\theta(\omega) - i\eta(\omega)\sin\theta(\omega)) \sim \theta(\omega) + i\eta(\omega)$, where $\theta$ is the rotation angle and $\eta$ is the ellipticity, for the small rotation angles. The magneto-optical Kerr measurements were performed with multiple reflections inside the substrate (InP). With respect to the waveform of the THz pulse transmitted through the sample (Fig. 2b), the second pulse comes from the multiple reflections inside the substrate after the first main pulse with a time delay $\Delta t = 2n_s d/c$, where $d$ is the thickness of the InP substrate ($d = 360$ μm), $c$ is the speed of light and $n_s$ is the refractive index of the InP substrate. By using $\Delta t$, $n_s$ is determined to be 3.46 which agrees with literature[39]. The second pulse experiences both the Faraday and Kerr effects when it is transmitted through the TI film and is reflected at the film-side surface of the substrate, respectively. Since the rotation angles of $\theta_F$ and $\theta_K$ are so small that the total rotation angle amounts to the sum of the Faraday and Kerr rotations, $\theta_K$ was obtained by subtracting the rotation angles for the first pulses from those of the second pulses. We have also confirmed that the Kerr rotations from the optical windows and the substrate under the magnetic fields are negligibly small (Extended Data Fig. 2b).

**Device fabrication.** For the accurate measurements of sheet resistance and Hall resistance[40,41], which are required for the accurate conversion to conductance, we designed a Hall bar device with (1) a well-defined aspect ratio of voltage probes, (2) point-like (i.e., thin) voltage probes so as to keep the equipotential lines uniform around them, and (3) the Hall probes well separated from the source and drain (typically, longer than the width) in order to avoid non-uniform equipotential lines around the source and drain under large Hall-angle conditions[31,40,41]. Thus, our Hall bars ($W$ = 200 μm in width and the 750-μm-long source-drain distance unless specified otherwise) have four voltage probes separated by $L$ = 200 μm as shown in Extended Data Fig. 3, in which the films were patterned by using photolithography and chemical wet etching with $H_2O_2$-$H_3PO_4$-$H_2O$ and $HCl$-$H_2O$ solutions. Electrical contact was made of Ti (5 nm)/Au (45 nm) by electron-beam evaporation. For gating the Hall bar devices, we deposited $AlO_x$ (~30 nm) by ALD and subsequently, Ti (5 nm)/Au (45 nm) electrodes were deposited by electron-beam evaporation. The films were patterned into Hall bars ($W$ = 200 μm in width and 600 μm in length unless specified otherwise) with four voltage terminals separated by $L$ = 200 μm (the inset of Fig. 3d) by using photolithography and chemical wet etching by $H_2O_2$-$H_3PO_4$-$H_2O$ and $HCl$-$H_2O$ solutions. Electrical contact was made of Ti(5 nm)/Au(45 nm) by electron-beam evaporation. For gating the Hall bar devices, we deposited $AlO_x$ (~30 nm) by ALD and subsequently, Ti(5 nm)/Au(45 nm) electrodes were deposited by electron-beam evaporation.

**Transport measurements.** Electrical transport measurements were carried out in a Quantum Design PPMS with a $^3$He probe in a variable temperature range of 0.5 to 300 K or with an adiabatic demagnetization refrigerator with a base temperature of 0.1 K, or in a dilution refrigerator with a base temperature of 40 mK. The electrical resistance was measured with the PPMS or with a standard lock-in technique. The excitation current was 10 to 100 nA to minimize the heating and the frequency was 3 Hz. The sheet resistance and the Hall resistance are derived by $\rho_{xx} = R_{xx}/(L/W)$ and $\rho_{yx} = R_{yx}$, respectively, where $L = W$ in the present samples (see Extended Data Fig. 3), and $R_{xx} = V_x/I_x$ and $R_{yx} = V_y/I_x$ ($V_{x,y}$ is the longitudinal, Hall voltage and $I_x$ is the excitation current) The sheet conductance $\sigma_{xx}$ and the Hall conductance $\sigma_{xy}$ were converted from $\rho_{xx}$ and $\rho_{yx}$ using the tensor relations: $\sigma_{xx} = \rho_{xx}/(\rho_{xx}^2+\rho_{yx}^2)$ and $\sigma_{xy} = \rho_{yx}/(\rho_{xx}^2+\rho_{yx}^2)$.

**Acknowledgments**

We thank J. G. Checkelsky for enlightening discussions, and K. N. Okada, S. Iguchi, M. Ogino, Y. Hayashi, H. Shishikura, D. Murata, and Y. Kato for support of the THz measurements. This research project was partly supported by JSPS/MEXT Grant-in-Aid for Scientific Research (No. 15H05853, 15H05867, 17J03179, 18H04229, 18H01155), and JST CREST (No. JPMJCR16F1, JPMJCR1874).

**Author contributions**

Y. Tokura conceived and supervised the project. M.M., R.Y., and K.Y. fabricated the samples with the help of A.T., K.S.T., and M. Kawasaki. M.M., Y.O., and Y. Takahashi performed the THz spectroscopy measurements and analyzed the data. M.M. and M. Kawamura performed the transport measurements and analyzed the data. T.M. and N.N. contributed to the theoretical discussions. M.M., M. Kawamura, T.M., N.N., and Y. Tokura wrote the manuscript with inputs from all the other authors.

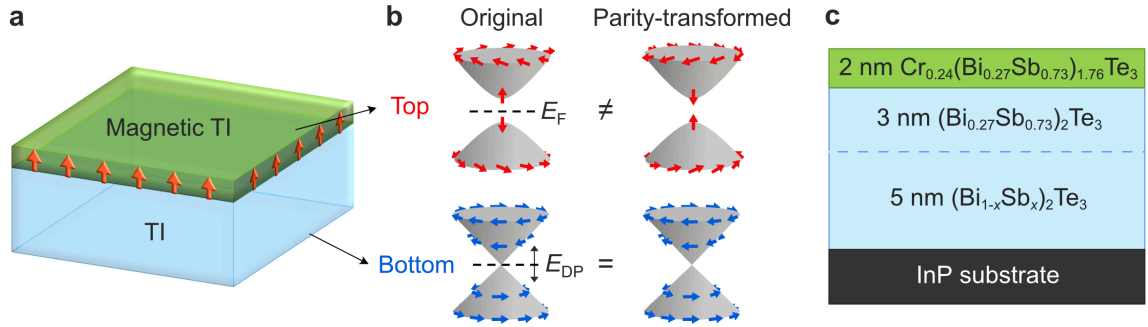

**Fig. 1 | Semi-magnetic topological insulator and the parity anomaly. a,b**, Schematic illustration of a semi-magnetic TI, where the top surface Dirac state is gapped while the bottom surface Dirac state is gapless. Orange arrows and red (blue) arrows indicate the magnetization and the spins for the top (bottom) surface Dirac cone, respectively. The 2D parity symmetry (i.e., mirror symmetry) of the gapped surface state is broken in that the spin-direction is not conserved by the parity operation. **c**, Schematic layout of a MBE-grown semi-magnetic TI film used for the experiments. The total thickness of 10 nm makes the hybridization between the top and bottom surface states negligible. In the upper part (5 nm from the top surface), the Bi:Sb ratio is fixed so that $E_F$ is located in the magnetically induced gap, whereas, in the lower part, the energy level of the Dirac point ($E_{DP}$) can be tuned by varying Sb fraction, $x$.

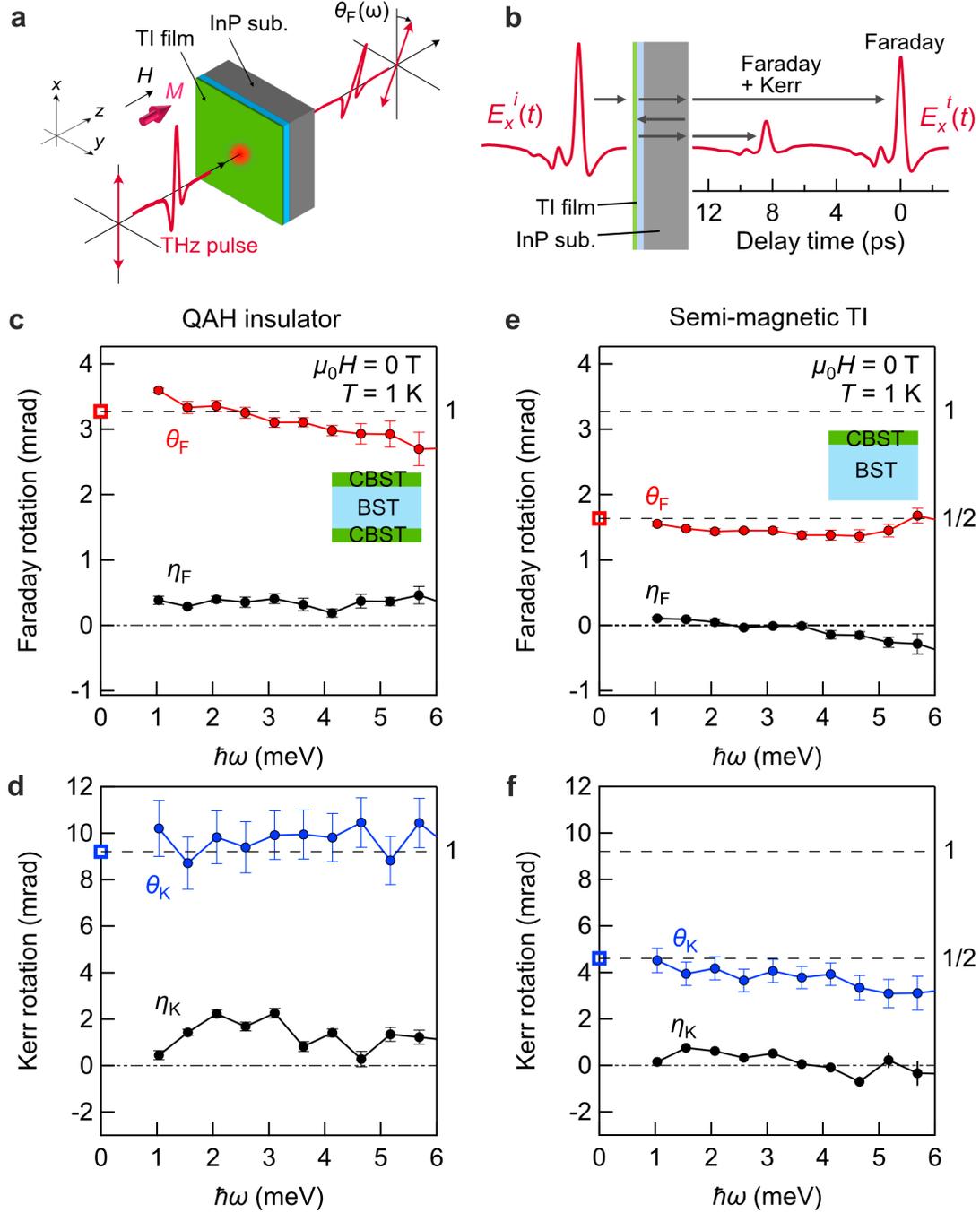

**Fig. 2 | Half quantized THz Faraday and Kerr rotations. a**, Schematic setup for the THz magneto-optical measurement for a TI thin film. **b**, Typical waveform of the incident and transmitted THz pulse. The waveform of the transmitted electric field is divided into the first and second transmitted pulses with a time delay due to multiple reflections in the substrate. **c-f**, Complex Faraday ($\theta_F + i\eta_F$) and Kerr ($\theta_K + i\eta_K$) rotation spectra for a QAH insulator film based on a magnetic TI heterostructure (see Supplementary Fig. S2) (**c** and **d**) and for a semi-

magnetic TI film ($x = 0.89$) (**e** and **f**) at $T = 1$ K and $\mu_0 H = 0$ T. The real ($\theta_F$, $\theta_K$) and imaginary ($\eta_F$, $\eta_K$) parts represent the polarization rotation angles and the ellipticities of the transmitted light, respectively. The error bars in (**c-f**) represent the standard error of the mean. The dashed lines and the open squares indicate the expected integer-quantized ($\theta_F = \tan^{-1}\frac{2\alpha}{n_s+1} = 3.26$ mrad and $\theta_K = \tan^{-1}\frac{4n_s\alpha}{n_s^2-1} = 9.21$ mrad) (**c** and **d**) and half-quantized ($\theta_F = \tan^{-1}\frac{\alpha}{n_s+1} = 1.63$ mrad and $\theta_K = \tan^{-1}\frac{2n_s\alpha}{n_s^2-1} = 4.60$ mrad) (**e** and **f**) rotation angles. BST and CBST, as schematically shown in (**c** and **e**), denote (Bi, Sb)$_2$Te$_3$ and Cr-doped (Bi, Sb)$_2$Te$_3$, respectively.

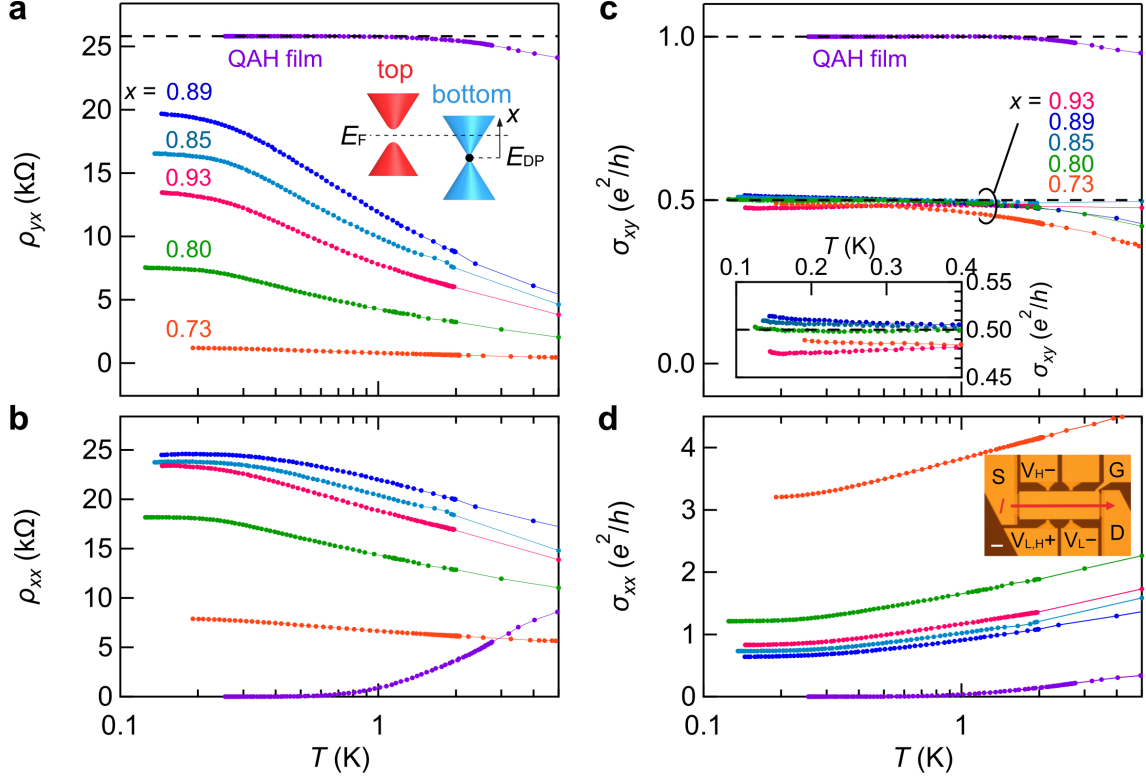

**Fig. 3 | Half-integer quantization in electrical transport. a-d**, Zero-field Hall resistance $\rho_{yx}$ (**a**), sheet resistance $\rho_{xx}$ (**b**), Hall conductance $\sigma_{xy}$ [$= \rho_{yx} / (\rho_{xx}^2 + \rho_{yx}^2)$] (**c**), and sheet conductance $\sigma_{xx}$ [$= \rho_{xx} / (\rho_{xx}^2 + \rho_{yx}^2)$] (**d**) as a function of logarithmic scale of temperature $T$. The data for the semi-magnetic TI films with various values of Bi:Sb ratio $x = 0.73, 0.80, 0.85, 0.89, 0.93$ and a typical QAH film sample (purple) are shown. The resistance measurement was conducted using Hall-bar-shaped samples as shown in the inset of (**d**): an optical microscope image of a Hall bar with a top-gate electrode. The white scale bar corresponds to 100 μm. The inset of (**a**) shows the schematic of the surface band structures for the semi-magnetic TI film, in which the carrier type is electrons (holes) when $x \leq 0.85$ ($\geq 0.89$) (Supplementary Figs. S8 and S9). The inset of (**c**) shows the magnified view of $\sigma_{xy}$ as a function of a linear scale of $T$.

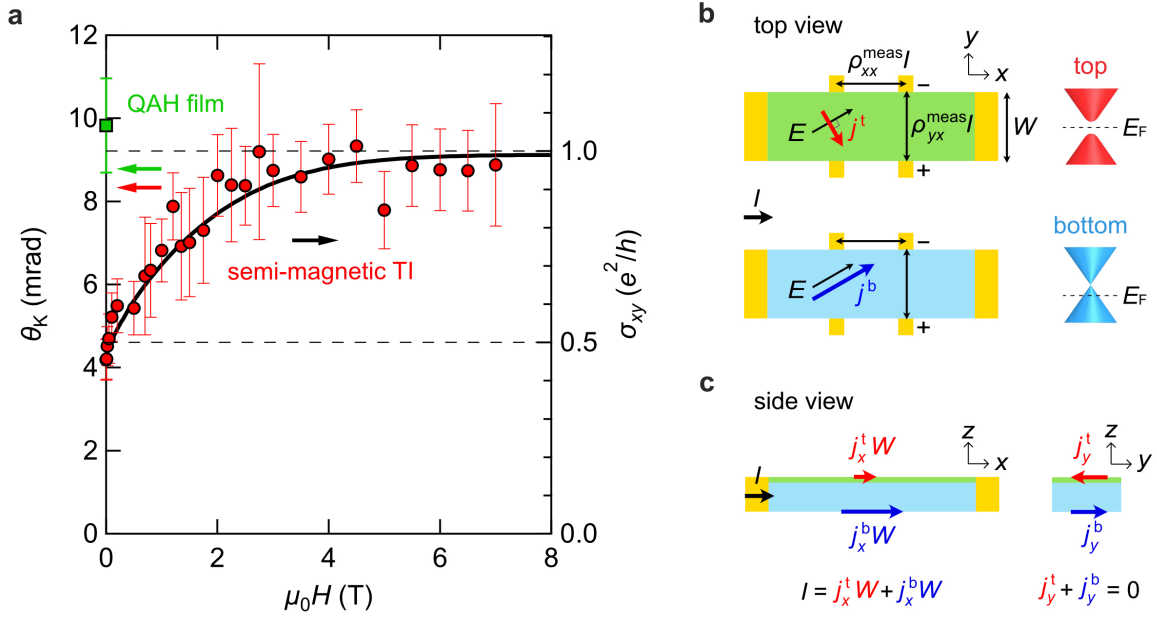

**Fig. 4 | Correspondence between transport and magneto-optics. a**, Magnetic field $\mu_0 H$ dependence of $\theta_K$ at $\hbar\omega = 2$ meV (red circles) and $\sigma_{xy}$ (black line) measured by dc transport at $T = 1$ K. The green square point represents $\theta_K$ for the QAH film at $\hbar\omega = 2$ meV. The ticks of the left and right ordinates are adjusted by using the relation of $\tan\theta_K = \frac{1}{c\varepsilon_0}\frac{2n_s}{n_s^2-1}\sigma_{xy}$. The error bars represent the standard error of the mean. **b**, **c**, Schematics of the top view (**b**) and the side view (**c**) of the current flow in the parallel conduction picture. The magnetic (top) and non-magnetic (bottom) layers are drawn in green and blue colors, respectively.

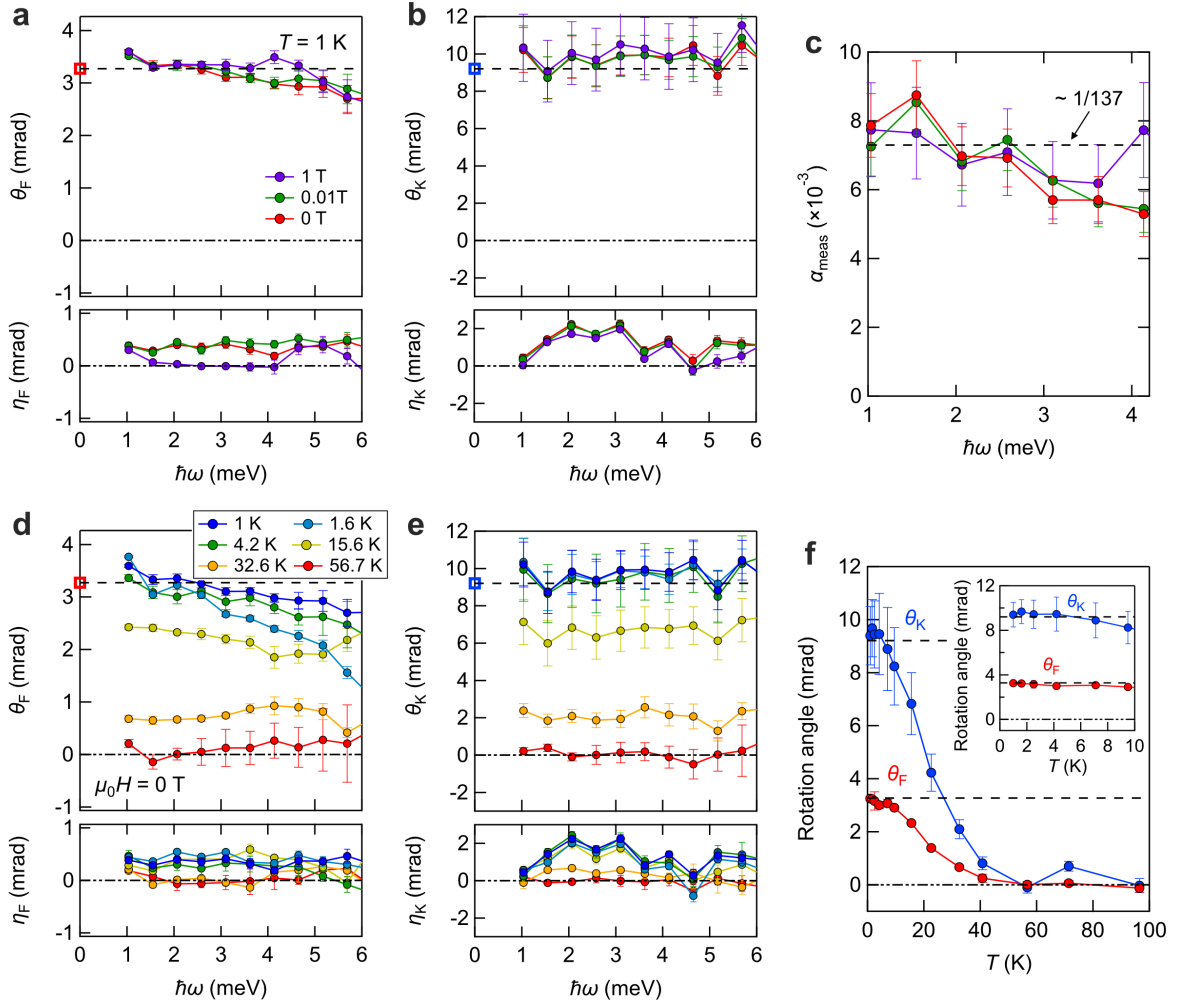

**Extended Data Fig. 1 | Quantized Faraday and Kerr rotations in a QAH state. a,b**, $\theta_F$ and $\eta_F$ (**a**) and $\theta_K$ and $\eta_K$ (**b**) spectra at $T = 1$ K for the QAH insulator film, which is partly the same as Figs. 2c and 2d in the main text, with a slight variation of external magnetic fields ($\mu_0 H = 0$, 0.01, and 1 T). **c**, Measured fine-structure constant $\alpha_{meas}$ which is calculated from (**a**) and (**b**) by using a relation of $\alpha_{meas} = (\tan\theta_F \tan\theta_K - \tan^2\theta_F)/(\tan\theta_K - 2\tan\theta_F)$ (18,24,25). **d, e**, $\theta_F$, $\eta_F$ (**d**) and $\theta_K$, $\eta_K$ (**e**) spectra at $\mu_0 H = 0$ T and at various temperatures ($T = 1$, 1.6, 4.2, 15.6, 32.6, and 56.7 K). **f**, $T$ dependence of $\theta_F$ and $\theta_K$ taken at $\hbar\omega = 2$ meV, suggesting that the Curie temperature is about 50 K and that the integer quantization subsists possibly up to 4.2 K. The inset is the magnified view of **f**. The error bars in a-f represent the standard error of the mean.

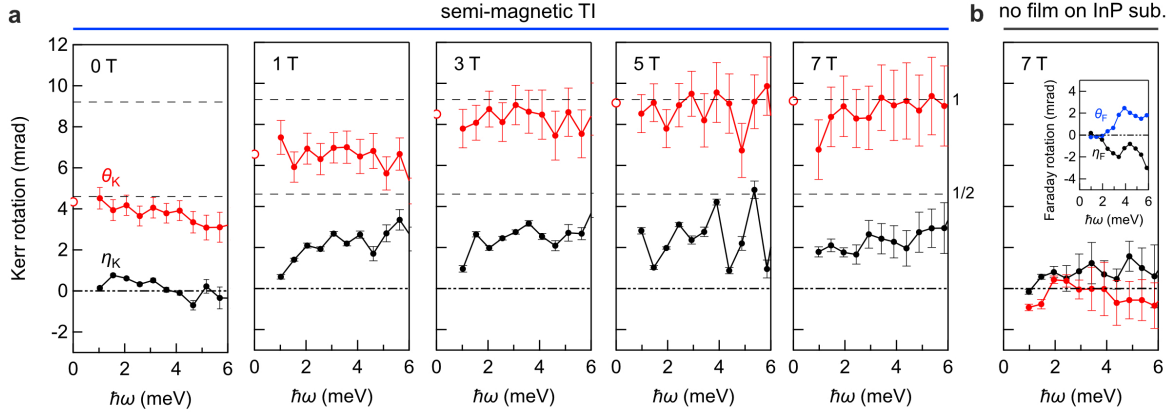

**Extended Data Fig. 2 | Kerr rotation in the semi-magnetic TI under magnetic fields. a,** Representative complex Kerr rotation spectra for the semi-magnetic TI film used for Fig. 4a in the main text. **b,** Background Kerr spectra of the InP substrate without any TI films at 7 T. The inset shows the Faraday rotation spectra at 7 T, where an observable polarization rotation occurs at $\hbar\omega > 2$ meV, possibly due to the magnetic resonance of magnetic impurities involved in InP substrates. The error bars in **a** and **b** represent the standard error of the mean.

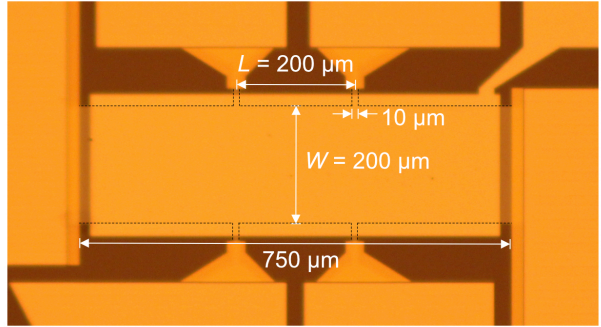

**Extended Data Fig. 3 | Optical microscope image of a typical Hall bar device used in the transport measurements.** The black broken lines indicate the shape of the TI film below the gate electrode, formed into the Hall bar structure.